\documentstyle[floats, aps, prb, epsfig]{revtex}

\draft

\begin{document}

\twocolumn[\hsize\textwidth\columnwidth\hsize\csname @twocolumnfalse\endcsname

\title{Thermodynamic Signature of a Two-Dimensional Metal-Insulator Transition }

\author{ S.~C.~Dultz and H.~W.~Jiang } \address{Department of Physics and
  Astronomy, University of California at Los Angeles,
  Los~Angeles,~CA~90095}

\date{\today}

\maketitle

\begin{abstract}
We present a study of the compressibility, $\kappa$, of a two-dimensional hole system which exhibits a metal-insulator phase transition at zero magnetic field. It has been observed that ${d\kappa\over dp}$ changes sign at the critical density for the metal-insulator transition. Measurements also indicate that the insulating phase is incompressible for all values of $B$.  Finally, we show how the phase transition evolves as the magnetic field is varied and construct a phase diagram in the density-magnetic field plane for this system.
\end{abstract}

]

\narrowtext
Recently, we are seeing a growing body of experimental evidence supporting a metal-insulator quantum phase transition in a number of two-dimensional electron \cite{Kravchenko} and hole systems\cite{Coleridge} where coulomb interactions are strong and particle mobility is quite high.  These experiments are of interest because of the prevailing theory of non-interacting particle systems in two dimensions which states that only insulating behavior should be seen at all densities for even the smallest amount of disorder in the system. \cite{Abrahams}  In order to further understand the nature of the unusual phase transition, it is important to study the thermodynamic properties near the transition.  One particular question is whether there is any signature for the phase transition in a thermodynamic measurement.  Theoretically, within the framework of Fermi liquid, one does not expect any qualitative change in the thermodynamic properties.\cite {Lee}  On the other hand, recent theories for strong interacting systems \cite{Varma,Chakravarty} have predicted that there should be profound consequences in thermodynamic measurements.

In this paper, we address this issue by presenting a measurement of one of the fundamental thermodynamic quantities: the thermodynamic density of states (or equivalently, the compressibility) of a strongly interacting two-dimensional hole system (2DHS). We report evidence that the compressibility measurement indeed provides an unambiguous signature for the metal-insulator transition (MIT).  The insulating phase is incompressible.  Furthermore, we show that the phase transition at $B = 0$ is intimately related to the quantum Hall state to insulator transition for the lowest Landau level in a finite magnetic field.

Traditionally, to obtain the density of states (DOS) or the compressibility of a 2D electron system, capacitance between the 2D electrons and the gate is measured. \cite{Goodall,Smith,Kravchenko1,Mosser}  This capacitance can be modeled as the geometrical capacitance in series with a quantum capacitance.  The quantum capacitance per unit area $c_{q}$ is related to the DOS  $\big({dn\over d\mu}\big)$ by $c_{q}= e^{2}\big({dn\over d\mu}\big)$, or to $\kappa$ by $c_{q} = n^{2} e^{2}\kappa$ where $\mu$ is the chemical potential and $n$ is the carrier density.  One major drawback of this method is that, in the low magnetic field limit, the quantum capacitance is much larger than the geometric capacitance.  For two capacitors in series, small uncertainty in the geometric capacitance can lead to a large quantitative error (even a sign error) in the extracted $\kappa$.  In a pioneering experiment by Eisenstein et al.\cite{Eisenstein}, the penetration of the electric field through 2D electrons in one well was detected by the 2D electrons in the other well by using a double quantum well sample.  This penetration field measures the screening ability of the electrons which was shown to be inversely proportional to $\kappa$.  For the present study, we extend the field penetration method to a more conventional heterostructure with only a single layer of carriers.

\begin{figure}[!t]
\begin{center}
\epsfig{file = 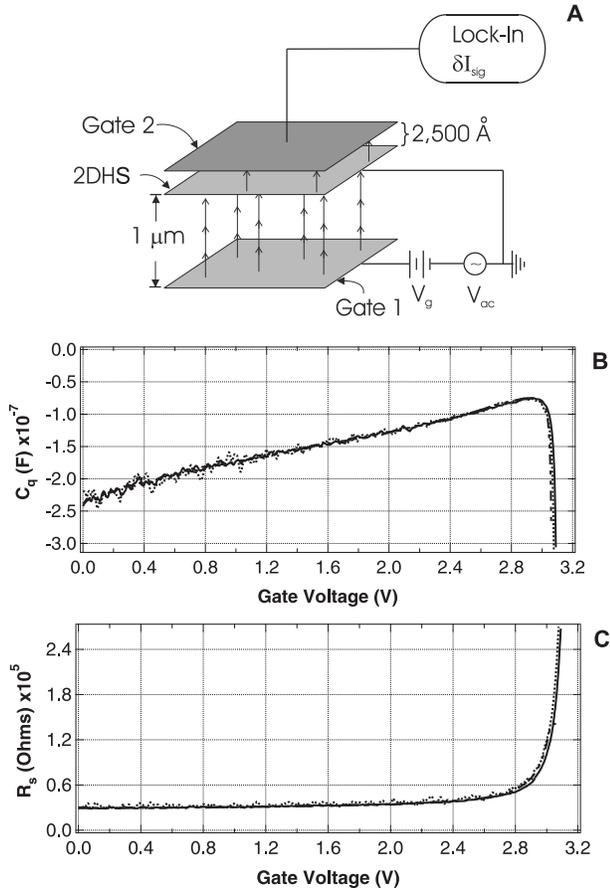, width = 8cm, clip=}
\end{center}
\caption{(a)  Diagram of our experimental setup with an active sample area of 1 $\text{mm}^{2}$.  (b) Typical trace of $C_{q}$ which is proportional to compressibility, and (c) $R_{s}$ which is a measure of dissipation as a function of the applied DC voltage at three different AC frequencies:  43.2 Hz, 100 Hz and 200 Hz.  All data was taken at $B=0$ and $T$ = 4.2 K.}
\label{Figure 1}
\end{figure}

The wafer used for the experimental devices was a p-type MBE grown GaAs/Al$_{x}$Ga$_{1-x}$As single heterostructure.  A $400~$\AA~Al$_{.45}$Ga$_{.55}$As undoped spacer was used to separate the Be-donors ($1~\times~10^{18}\text{~cm}^{-3}$)  from the 2DHS. Just below the 2DHS is a $5000~$\AA~undoped GaAs buffer layer and beneath that, a $5000~$\AA~Al$_{.72}$Ga$_{.28}$As layer which was used as the etching stopping layer for substrate removal. The mobility of this particular sample was roughly $123,000~\text{~cm}^2/\text{V-s}$ with a hole density of $p=2.60~\times~10^{11}\text{~cm}^{-2}$.  The device for the compressibility measurement was fabricated by sandwiching the 2DHS between two metallic electrodes.  To form the top electrode ($2500~$\AA~from the 2DHS), NiCr was evaporated on the surface of the sample.  To form the bottom electrode, the GaAs substrate was totally removed so that another NiCr electrode could be placed on the bottom in close proximity to the 2DHS ($10000~$\AA).  Details of the substrate removal can be found elsewhere \cite{Weckwerth}.  To measure the compressibility, we applied a 10 mV AC excitation voltage $V_{ac}$ to the bottom electrode (Gate 1), as shown in Fig. \ref{Figure 1}a.  A DC voltage $V_{g}$ was superimposed to vary the carrier density.  The 2DHS was grounded to screen the electric field from the bottom electrode.  A lock-in amplifier was used to detect the penetrating electric field as current from the top electrode (Gate 2) to ground.  By modeling the system as a distributed circuit, both the quantum capacitance $C_{q}$ and the resistance $R_{s}$ of the channel for the 2DHS could be extracted individually based on the measured values for the in-phase $I_{x}$ and $90^{o}$ phase $I_{y}$ current components.  The model for the circuit is an extension of the two wire transmission line problem \cite{Wangsness}.  The exact expression for current is as follows:
$$
I={i\omega C_{1}C_{2}V_{ac}\over C_{1}+C_{2}}
\bigg[1-\bigg({C_{q}\over C_{1}+C_{2}+C_{q}}
\bigg){\tanh (\alpha )\over \alpha }
\bigg]
;$$
$$
\alpha =\sqrt {i\omega {C_{q}(C_{1}+C_{2})\over C_{1}+C_{2}+C_{q}}R_{s}}
;$$
In this expression, $\omega$ is the frequency of our excitation voltage on gate 1 and $V_{ac}$ is the amplitude. $C_1$ (136 pF) and $C_2$ (541 pF) are the geometric
capacitances between gate 1 and gate 2 and the 2DHS, respectively.  This is a complex equation with the real and imaginary parts coupled with respect to $R_{s}$ and $C_{q}$.  These values were obtained by solving both equations simultaneously.\cite {Dultz1}  In the low-frequency limit, $I_{x}$ is directly proportional to $-R_{s}$, the dissipation of the 2DHS, and $I_{y}$ is proportional to $1/C_{q}$, the inverse compressibility.

Fig. \ref{Figure 1}b and \ref{Figure 1}c show typical traces of the $C_{q}$ and  $R_{s}$ as a function of the gate voltage at various different frequencies. It is apparent that there is no frequency dependence over the entire range of gate voltage.  Furthermore, we found that, for frequencies up to 200 Hz, $I_{x}$ and $I_{y}$ are indeed directly proportional to $-R_{s}$ and $1/C_{q}$ respectively.  It can then be assumed that the divergence in both channels is not due to depletion of the channel since this would produce a noticeable frequency dependence on the high gate voltage side of the maximum in $C_{q}$.

\begin{figure}[!b]
\begin{center}
\epsfig{file = 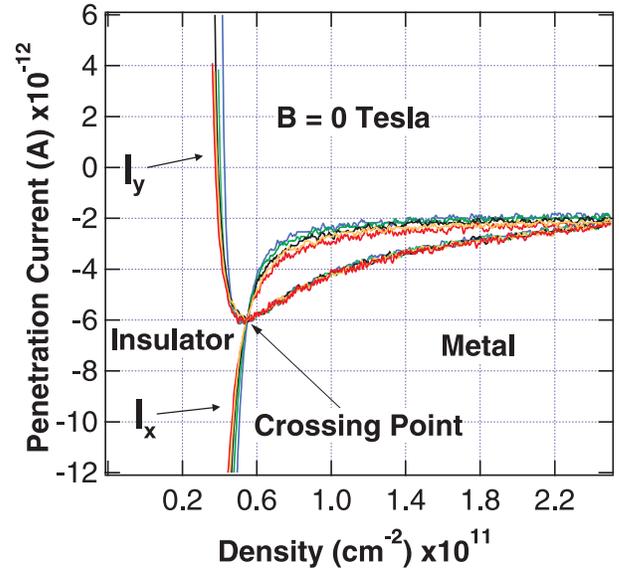, width = 8cm, clip=}
\end{center}
\caption{$I_{x}$ and $I_{y}$ vs density for five temperatures at an excitation frequency of 100 Hz:  blue- 0.33 K, green- 0.56 K, black- 0.82 K, orange- 1.02 K, red- 1.28 K.  The crossing point of the five dissipation channel curves corresponds to the metal-insulator phase transition at $B=0$.  The minimum in the inverse compressibility channel occurs at the same density of $p=5.5\times 10^{10}\text{~cm}^{-2}$.}
\label{Figure 2}
\end{figure}

Fig. \ref{Figure 2} shows both the inverse compressibility and the dissipation signals as a function of the density at $B=0$ for different temperatures ranging from 0.33 K to 1.28 K.  Two main features are immediately noticeable for the $1/\kappa$ channel.  First, $1/\kappa$ is negative for high densities and becomes more negative as the density is decreased as others have seen in diluted electron \cite{Eisenstein} and hole systems\cite{Shapira}. The negative compressibility is known to be due to the strong exchange energy contribution to the total chemical potential\cite{Eisenstein1}.  Secondly, a sharp turn-around occurs at $p=5.5\times 10^{10}\text {~cm}^{-2}$ as ${d\kappa\over dp}$ changes sign. As the density is further reduced, $1/\kappa$ becomes positive and diverges rapidly (extraction of $C_{q}$ shows that $\kappa$ rapidly approaches zero from infinity in the low density limit \cite {Dultz1}).  Focusing now on the dissipation channel, one can find a temperature independent crossing point which we believe to be the critical density for the MIT in this sample (also can be seen in transport measurements).  Although the qualitative shape of $I_{y}$ as a function of gate voltage was seen by others \cite{Eisenstein} the minimum in this signal was never recognized as the MIT.  Notice the temperature dependence on the high density side of the crossing point.  As the temperature increases, $I_{x}$ is getting more negative which means that the 2DHS is getting more resistive (metallic behavior).  The opposite is seen on the low density side of this crossing point where the characteristic temperature dependence is that of an insulator.  This crossing point occurs at the point where ${\partial \kappa\over \partial p}$ changes sign precisely at the minimum of $1/\kappa$.  Therefore, we believe there is a clear signature of the metal-insulator phase transition at $B=0$ in this thermodynamic measurement.  Since $\kappa$ tends toward zero in the insulating phase, the data also suggests that the insulating phase is incompressible.  Theoretically, it has been argued that the insulating phase is a Wigner glass phase which is incompressible for strongly interacting systems.\cite{Varma,Chakravarty}

\begin{figure}[!b]
\begin{center}
\epsfig{file = 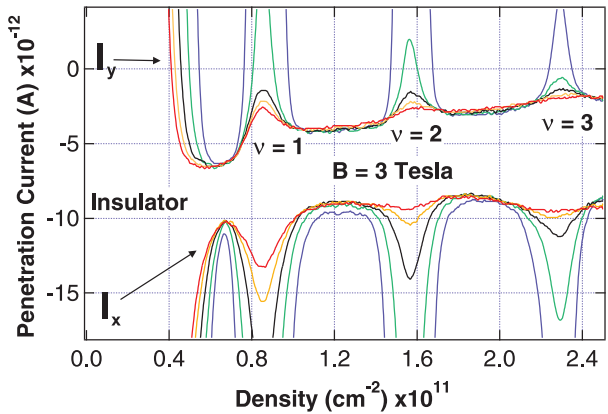, width = 8cm, clip=}
\end{center}
\caption{$I_{x}$ and $I_{y}$ vs density for five temperatures at $B$ = 3 T and excitation frequency, 100 Hz:  blue- 0.33 K, green- 0.56 K, black- 0.81 K, orange- 1.05 K, red- 1.27 K.  The crossing point now becomes the DOS peak of the first Landau level.  Peaks in the $I_{y}$ channel represent regions where the DOS is going to zero.  The $I_{x}$ channel corresponds to the set of lower curves.}
\label{Figure 3}
\end{figure}

Having identified that the signature of the MIT has been seen in $\kappa$ at $B=0$, we would like to see how this critical point evolves as the magnetic field is increased.  We found $1/\kappa$ vs. $p$ to be independent of magnetic field up to about 1.5 T. Fig. \ref{Figure 3} shows how $1/\kappa$ and $R_{s}$ evolve as a function of density in a higher magnetic field where variations in both $I_{x}$ and $I_{y}$ are more dramatic due to the presence of Landau levels.  The data shown was taken at 3 T and at five different temperatures from 0.33 K to 1.27 K just as in the $B=0$ case.  As seen in the figure, $I_{y}$ shows a local maximum in an integer filling factor where the compressibility, which is also proportional to the DOS, tends to zero between two adjacent Landau levels. The DOS is zero only at $T$ = 0, so the peaks get more pronounced as one goes to lower temperatures.  Conversely, $I_{y}$ reaches a minimum when the DOS reaches a peak in a Landau level center where the delocalized states (states which exist at the Landau level centers where the electronic wavefunction is expected to extend spatially throughout the sample) reside.  Meanwhile, $I_{x}$ also undergoes oscillations. It is important to note here, $I_{x}$ in the high field is not proportional to $\rho_{xx}$ but rather to a constant term plus a term that goes like 1/$\sigma_{xx}$ in the limit of high conductivity \cite{Goodall}.  Because $\sigma_{xx}$ is also going to zero in the quantum Hall liquid regime, this gives us a peak in $I_{x}$ when the DOS tends to zero. At 3 T, we also see a temperature independent point in both channels and in the same place where the inverse compressibility reaches a minimum.  In this case however, the temperature independent point marks the phase boundary between the insulator and the $\nu=1$ quantum Hall state.  On the insulator side, $1/\kappa$ diverges just as in the $B=0$ case and so one cannot distinguish between the insulating phase at $B=0$ versus the insulator at finite field.

\begin{figure}[!b]
\begin{center}
\epsfig{file = 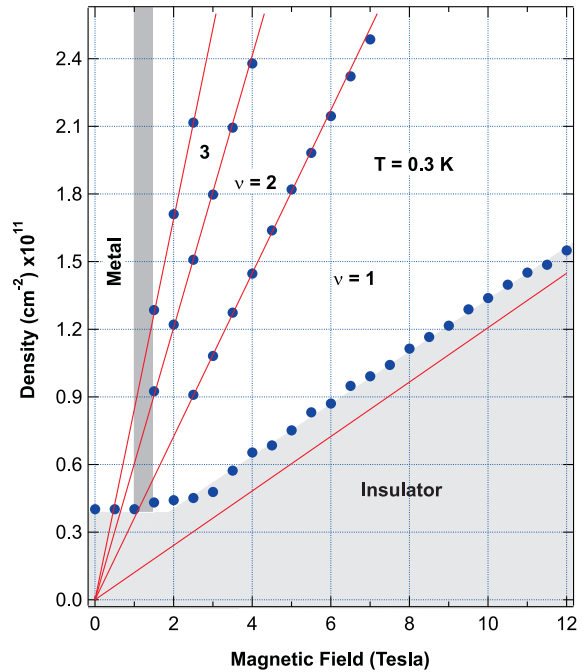, width = 7.5cm, clip=}
\end{center}
\caption{A map of the phase diagram in the $p-B$ plane for the 2DHS at T = 0.3 K. The dots are data taken from graphs like figure \ref{Figure 3} where the local minimums exist in $I_{y}$ (peak in DOS) and the temperature independent crossing points exist in $I_{x}$. The uncertainty in peak position is approximately $5\times 10^{9}\text {~ cm}^{-2}$ in density.  The lines are calculated positions for Landau level centers in the density-magnetic field plane from standard non-interacting particle picture.}
\label{Figure 4}
\end{figure}

As we vary the magnetic field from 0 to 12 T, we keep track of where the $1/\kappa$ minimums (i.e. phase boundaries) are occurring in density.  Fig. \ref{Figure 4} is a phase diagram in the density-magnetic field plane.  We can see how the DOS peaks are evolving as one goes from the high field to the low field regime.  There are number of interesting features.  First, the phase boundary for the lowest Landau level flattens out as $B$ is reduced.  If we can assume that the delocalized states are occurring in the DOS maximum, we can draw the conclusion that the delocalized states do not float up as $B$ tends toward zero for the strongly interacting 2DHS.  The $r_{s}$ value which is the ratio of the Coulomb energy to the Fermi energy, reaches about 20 near the MIT.  For non-interacting electron systems with low $r_{s}$ values, the lowest delocalized states are found to ``float up" in energy\cite{Glozman} as $B\rightarrow0$.  This energy divergence means that the Fermi energy can only be tuned through localized states which do not contribute to current but give only insulating behavior at $B=0$ (i.e.- resistivity diverges as temperature decreases).  The current observation is consistent with the studies that have been done in the past for the 2DHS through transport measurements. \cite{Dultz,Shahar} This implies that for the 2DHS there is indeed a metallic regime (as shown) in the thermodynamic limit which does not exist in the phase diagrams for the 2DES.  We would like to note here the data in Fig. \ref{Figure 4} was taken from another 2DHS sample which was cut from the same wafer but had slightly lower density and we see the lowest delocalized state terminating at a density of roughly $4.0\times 10^{10}\text{~ cm}^{-2}$ at $B=0$ rather than a density of $5.5\times 10^{10}\text{~cm}^{-2}$ as seen in Fig. \ref{Figure 2}.  Although we have only clearly seen the phase boundary between the metallic and insulating states in low magnetic field as we cross the critical density, we can see from our data that the metallic phase exists all the way to our highest density of $2.6\times 10^{11}\text{~cm}^{-2}$ at low magnetic fields.  We have also seen that this metallic phase is preserved to at least 1 T in our sample.  The shaded vertical band marks an ill-defined region between the metallic phase and the quantum Hall phases.  This region needs to be explored in more detail at lower temperatures when the quantum Hall plateaus are better resolved.  The insulating regime is also shaded in the diagram for clarity.  Secondly, the data suggests that the insulator to $\nu=1$ transition is related to the $B=0$ MIT based upon the similar behavior of the compressibility.  In fact, the same argument has been made based on the tracking of temperature independent crossing points in a transport experiment.\cite{Shahar}

In summary we have found, using an improved electric field penetration technique, the compressibility of a strongly interacting hole system undergoes a qualitative change traversing the MIT. More importantly, we have observed that the local maximum in compressibility occurs precisely at the critical density for the MIT at $B=0$.  The divergence of the compressibility shows the insulating phase is incompressible for all values of magnetic field.  The phase transition at $B=0$, in fact,  evolves into the quantum Hall to insulator transition at high $B$.  We believe these observations reported here cannot be explained by simple non-interacting models.  Theoretical analysis for this strongly interacting system is called for to understand the thermodynamic properties presented here.

The authors would like to thank S. Chakravarty, Q. Shi, J. Simmons, S. Sondhi, and C. Varma for helpful discussions, and B. Alavi for technical assistance. This work is supported by NSF under grant \# DMR 9705439.


\begin{references}
\bibitem {Dultz1}
detailed procedures and results, particularly in the insulating phase, will be presented in a longer publication.
\end{references}
\end{document}